\documentclass[aps,Phys. Rev. R,a4paper,twocolumn,superscriptaddress,citeautoscript]{revtex4-1}
\usepackage[english]{babel}
\usepackage[T1]{fontenc}
\usepackage{amsmath}
\usepackage{hyperref}
\usepackage{units}
\usepackage[charter]{mathdesign}
\usepackage[pdftex]{graphicx}
\usepackage{color}
\begin{document}
\title{Unconventional free charge in the correlated semimetal Nd$_2$Ir$_2$O$_{7}$}
\author{K. Wang}
\affiliation{Department of Quantum Matter Physics, University of Geneva, 24 Quai Ernest-Ansermet, 1211 Geneva 4, Switzerland}
\author{B. Xu}
\affiliation{Department of Physics and Fribourg Center for Nanomaterials, University of Fribourg, Chemin du Mus\'{e}e 3, CH-1700 Fribourg, Switzerland}
\author{C. W. Rischau}
\affiliation{Department of Quantum Matter Physics, University of Geneva, 24 Quai Ernest-Ansermet, 1211 Geneva 4, Switzerland}
\author{N. Bachar} 
\affiliation{Department of Quantum Matter Physics, University of Geneva, 24 Quai Ernest-Ansermet, 1211 Geneva 4, Switzerland}
\author{B. Michon}
\affiliation{Department of Quantum Matter Physics, University of Geneva, 24 Quai Ernest-Ansermet, 1211 Geneva 4, Switzerland}
\author{J. Teyssier}
\affiliation{Department of Quantum Matter Physics, University of Geneva, 24 Quai Ernest-Ansermet, 1211 Geneva 4, Switzerland}
\author{Y. Qiu}
\affiliation{Institute for Solid State Physics, The University of Tokyo, Kashiwa 277-8581, Japan}
\author{T. Ohtsuki}
\affiliation{Institute for Solid State Physics, The University of Tokyo, Kashiwa 277-8581, Japan}
\author{Bing Cheng}
\affiliation{Institute for Solid State Physics, The University of Tokyo, Kashiwa 277-8581, Japan}
\author{N.P. Armitage}
\affiliation{The Institute for Quantum Matter and the Department of Physics and Astronomy, The Johns Hopkins University, Baltimore, MD 21218, USA}
\author{S. Nakatsuji}
\affiliation{Institute for Solid State Physics, The University of Tokyo, Kashiwa 277-8581, Japan}
\affiliation{The Institute for Quantum Matter and the Department of Physics and Astronomy, The Johns Hopkins University, Baltimore, MD 21218, USA}
\affiliation{CREST, Japan Science and Technology Agency, Kawaguchi, Saitama 332-0012, Japan}
\affiliation{Department of Physics, University of Tokyo, Hongo, Bunkyo-ku, Tokyo 113-0033, Japan}
\author{D. van der Marel}\email{dirk.vandermarel@unige.ch}
\affiliation{Department of Quantum Matter Physics, University of Geneva, 24 Quai Ernest-Ansermet, 1211 Geneva 4, Switzerland}
\date{\today}
\%
\begin{abstract}
Nd$_2$Ir$_2$O$_{7}$ is a correlated semimetal with the pyrochlore structure, in which competing spin-orbit coupling and electron-electron interactions are believed to induce a time-reversal symmetry broken Weyl semimetal phase characterized by pairs of topologically protected Dirac points at the Fermi energy~\cite{wan2011,tian2016,armitage2018,ohtsuki2019}.   
However, the emergent properties in these materials are far from clear, and exotic new states of matter have been conjectured~\cite{pesin2010,moon2013,morimoto2016}.
Here we demonstrate optically that at low temperatures the free carrier spectral weight is proportional to $T^2$ where $T$ is the temperature, as expected for massless Dirac electrons. However, we do {\em not} observe the corresponding $T^3$ term in the specific heat. That the system is not in a Fermi liquid state is further corroborated by the "Planckian"~\cite{zaanen2019} $T$-linear temperature dependence of the momentum relaxation rate and the progressive opening of a correlation-induced gap at low temperatures.
These observations can not be reconciled within the framework of band theory of electron-like quasiparticles and point toward the effective decoupling of the charge transport from the single particle sector.
\end{abstract}
\maketitle

\section{Introduction}
Topological metallic states in correlated systems with strong spin-orbit coupling are an active field of research, which has in recent years lead to the observation of a topological Kondo insulator in SmB$_6$~\cite{dzero2010,kim2014}, a magnetic Weyl semimetal in Mn$_3$Sn~\cite{nakatsuji2015,kuroda2017} and a Weyl Kondo phase in Ce$_3$Bi$_4$Pd$_3$~\cite{lai2018} and CeRu$_4$Sn$_6$~\cite{xu2017}. Of particular interest in this context are the transition metal oxides with the pyrochlore structure, having the composition Ln$_2$Ir$_2$O$_{7}$ where Ln is usually a trivalent rare earth ion. The primitive cell contains two formula units, {\em i.e.} 4 Ir ions, with the Ir-O-Ir bonds forming a frustrated network of corner-sharing tetrahedra. 
At high temperature these materials are paramagnetic semimetals where a pair of doubly degenerate bands of opposite curvature have a quadratic band touching (QBT) point at the center of the Brillouin zone~\cite{kondo2015,cheng2017,moon2013} that at zero doping is coincident with the Fermi level and the Fermi surface is a single point.
At low temperature these materials were predicted~\cite{wan2011} and observed ~\cite{tomiyasu2012,guo2013,sagayama2013,ma2015,donnerer2016,chun2018} to order magnetically with all the Ir $5d$ moments pointing inward or outward from the center of the tetrahedron. 
This order does not affect the translational symmetry of the system, but since it breaks the time reversal symmetry, the double degeneracy of the bands is lifted, giving rise to 4 non-degenerate bands~\cite{wan2011}. 
Simultaneously the zone-center QBT transforms into 12 pairs of Weyl nodes, making for a grand total of 24 Weyl nodes~\cite{wan2011}. 
These nodes act as a source or sink of Berry curvature, and are topologically protected.
Increasing the on-site Coulomb interaction shifts the Weyl nodes away from the zone center towards the Brilloun zone boundary, until the two members of each pair meet at the zone boundary where they mutually annihilate and a correlation-induced gap opens~\cite{wan2011,go2012,witczak-krempa2012,witczak-krempa2013,shinaoka2015,morimoto2016,tian2016,ueda2017}. 
Ueda {\em et al.} have confirmed the presence of a gap at low temperatures in the optical spectrum of polycrystalline Nd$_2$Ir$_2$O$_7$~\cite{ueda2012,ueda2016}.
Indications of Weyl semimetal behavior have been reported in the optical spectra of the related compound Eu$_2$Ir$_2$O$_7$\cite{sushkov2015,ohtsuki2019,machida2010} at low temperatures. 
\begin{figure}[!t]
\begin{center}
\includegraphics[width=\columnwidth]{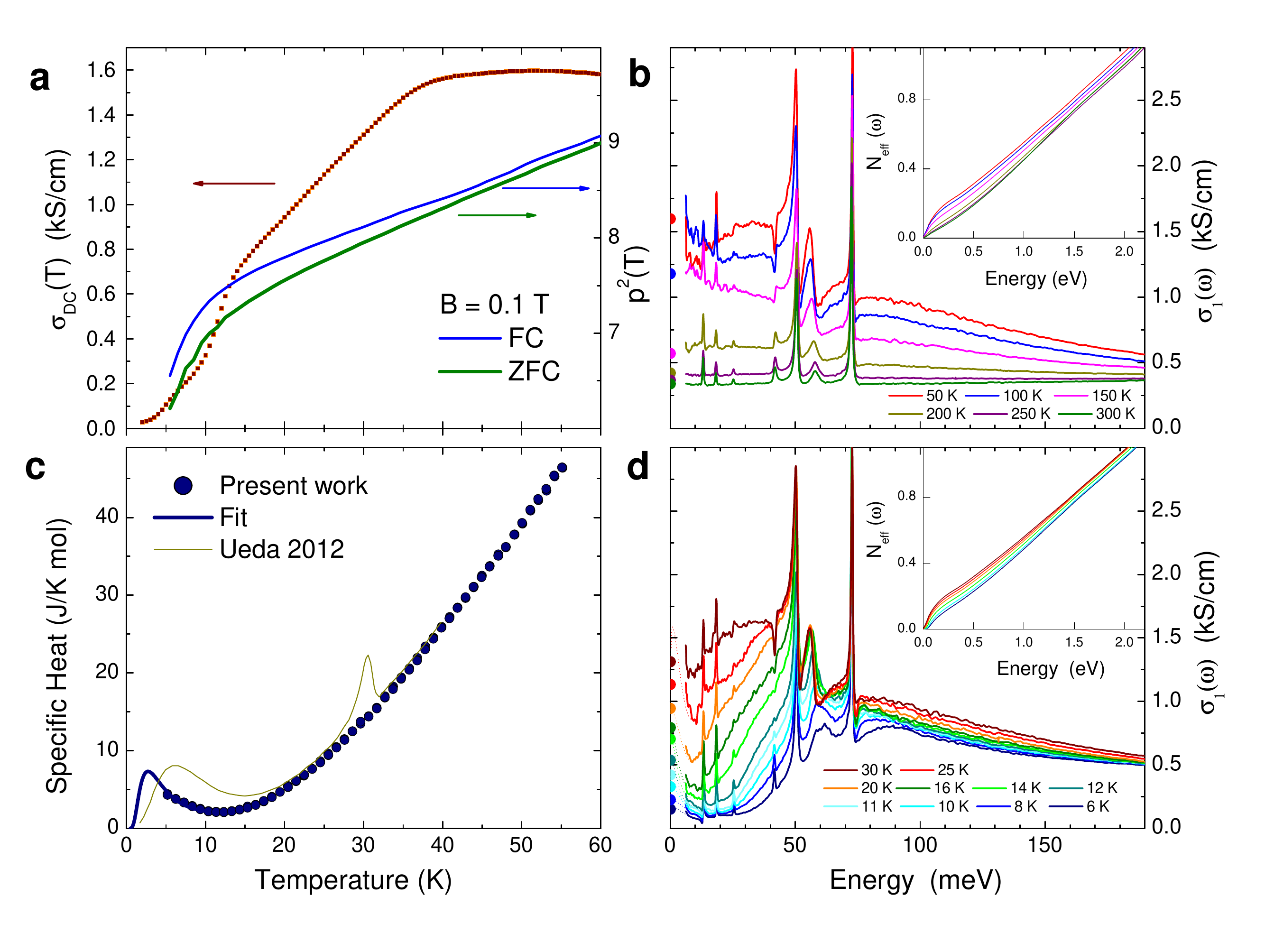}
\caption{ 
{\bf Experimental transport and optical conductivity data of Nd$_2$Ir$_2$O$_{7}$.} 
{\bf a}, DC conductivity (orange symbols). Blue and green curves are the ZFC and FC temperature dependence of the squared effective moment per Nd ion $p(T)^2 \equiv (V_{pc}/4)   \chi(T) 3 k_B T  / \mu_B^{2}$, where $k_B$ is the Boltzmann constant, $\mu_B$ the Bohr magneton,
$V_{pc}$ the volume of the primitive cell and $\chi(T)$ the magnetic susceptibility (Appendix, Eq.~\ref{eq:p}). If the Curie law applies one should expect $p(T)^2=13.1$ at all temperatures, corresponding to the local moment of a Nd$^{3+}$ ion. The decrease of $p(T)^2$ for decreasing temperature reveals the departure from the Curie-law as a result of exchange coupling of the magnetic ions. 
{\bf b}, Optical conductivity of Nd$_2$Ir$_2$O$_{7}$ at selected temperatures above 37~K. 
{\bf c}, Dark blue circles are the present specific heat data. The dark blue solid curve is a fit to Eq.~\ref{eq:schottky_c} (Appendix) with $\Delta_0=6.5$~K. The dark yellow curve is the specific heat of polycrystalline Nd$_2$Ir$_{2}$O$_{7}$ digitized from Fig.~1 of Ref.~\onlinecite{ueda2012}. 
{\bf d}, Optical conductivity of Nd$_2$Ir$_2$O$_{7}$ at selected temperatures below 37~K. Solid dots along the ordinates of {\bf b} and {\bf d}: DC conductivity using the 4 terminal method.
Insets: Effective number of electrons per primitive cell (Appendix, Eq.~\ref{eq:neff}).
}
\end{center}
\end{figure}
\section{Experimental results}
We have measured and analyzed transport, specific heat, magnetic susceptibility and optical spectroscopic data of single crystalline Nd$_2$Ir$_2$O$_{7}$. Experimental methods are described in the Appendix. In Fig.~1 the DC conductivity, susceptibility, specific heat, and low energy optical conductivity data are shown. In Fig.~2 the optical conductivity is shown on an expanded scale. 
Fitting the reflectivity and ellipsometry data to a Drude-Lorentz model provided the plasma-frequency $\omega_p(T)^2$ of the zero energy mode displayed in Fig.~2{\bf a}. The free carrier weight shows a $T^2$ temperature dependence with a small zero-temperature offset $\omega_p(0)^2$. 
It is expected for a system with $g$ Dirac cones at the Fermi level that~\cite{tabert2016a,tabert2016b}
\begin{equation}
\omega_p(T)^2=\omega_p(0)^2+ \frac{2\pi \alpha}{9} \frac{g c }{\upsilon_{}}  \left(k_BT\right)^2 
\nonumber
\end{equation}
where $\alpha$ is the fine structure constant, $c$ the light velocity in vacuum, and $k_B$ the Boltzmann constant.
Following Ref.~\onlinecite{wan2011} we will assume that there are 12 Weyl pairs, {\em i.e.} $g=24$. 
The experimental data in Fig.~2{\bf a} then imply that $\upsilon_{}=3.1$ km/s. The $\omega_p(0)^2$ term indicates the presence of doped charge carriers and, for the aforementioned values of $g$ and  $\upsilon$, corresponds to $0.003$ carriers per Ir ion and $|\mu|=1.0$~meV (Appendix, Eqs.~3-4). 
This small value of  $|\mu|$ may imply that, following the argument of Moon {\em et al.}, the Coulomb interaction can convert the paramagnetic QBT into a quantum critical non-Fermi-liquid~\cite{moon2013}.
The optical conductivity of polycrystalline Nd$_2$Ir$_2$O$_7$ shows a somewhat more pronounced gap at low temperatures and a smaller weight of the zero energy mode, but the temperature dependence of the spectral weight of the zero energy mode was not addressed~\cite{ueda2012}. 
\begin{figure}[!t]
\begin{center}
\includegraphics[width=\columnwidth]{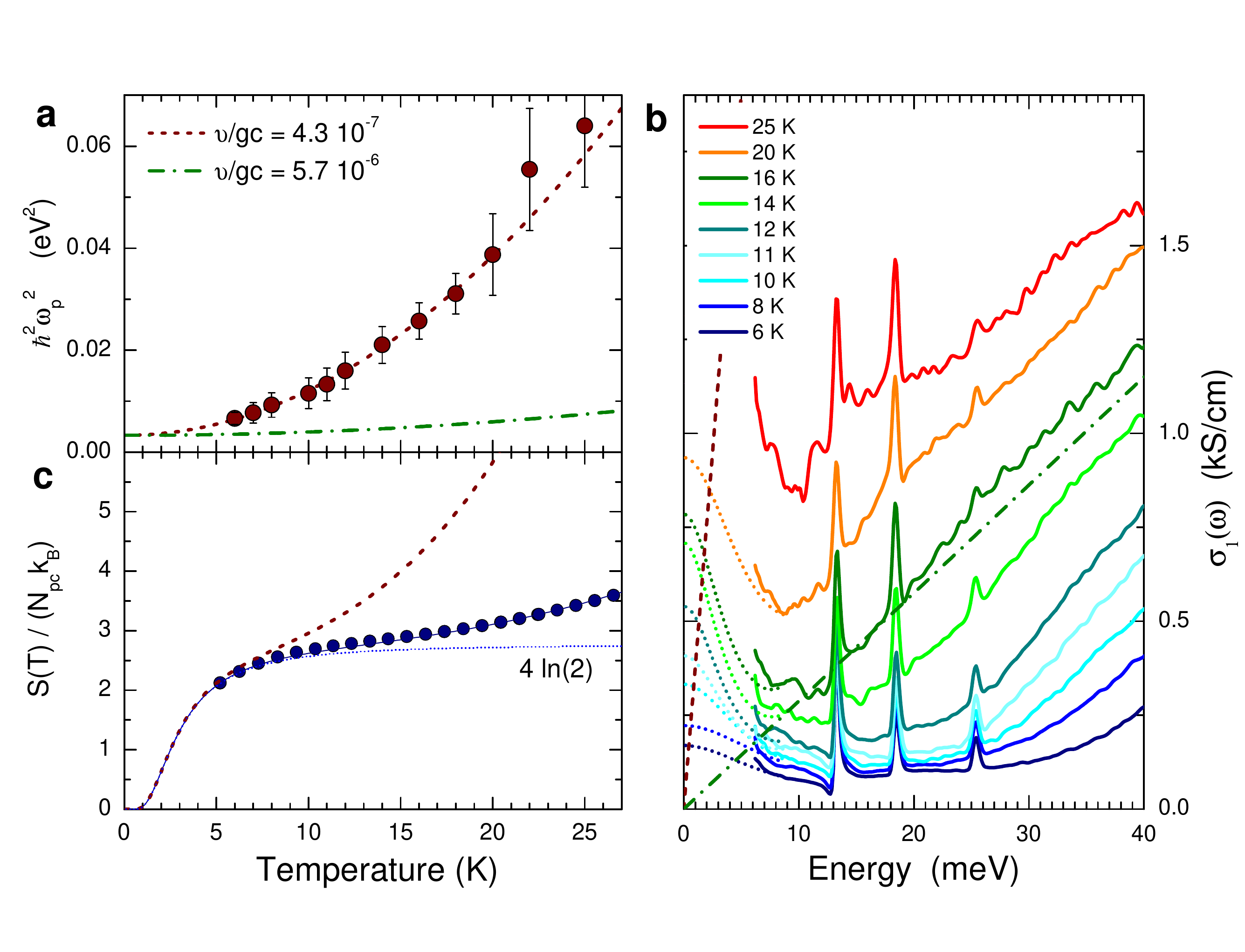}
\caption{ 
{\bf Comparison and analysis of free charge spectral weight, entropy and low energy optical conductivity of Nd$_2$Ir$_2$O$_{7}$.} 
{\bf a}, Brown dots: Temperature dependence of the experimentally obtained free carrier spectral weight. {\color{black} The error bars were obtained by shifting the reflectivity data $4\%$ up and down relative to the measured values.} Brown (green) dashed curve: Theoretical values (Appendix, Eq.~\ref{eq:cW}) for $g$ Weyl nodes in the first Brillouin zone with velocity $\upsilon_{}/gc=4.3\cdot 10^{-7}$ ($5.7\cdot 10^{-6}$), where $c$ is the light velocity in vacuum.
{\bf b}, Optical conductivity for selected temperatures. Dark green dash-dotted line: Theoretical $\sigma_1(\omega)$ (Appendix, Eq.~\ref{eq:sigslope}) for  $\upsilon_{}/gc=5.7\cdot 10^{-6}$.
{\bf c} Entropy per primitive cell (two formula units of Nd$_2$Ir$_2$O$_{7}$) in units of $k_B$. Dark blue curve: Integration of the fit to the experimental specific heat data of Fig.~1, using Eq.~\ref{eq:S} of the Appendix. Blue dots: integration of the experimental data of Fig.~1. Dotted blue line: Schottky contribution.
Brown dashed curve: Sum of the Schottky contribution and the theoretical prediction for Weyl fermions (Appendix, Eq.~\ref{eq:cW}) with velocity $\upsilon_{}/gc=4.3\cdot 10^{-7}$. 
}
\end{center}
\end{figure}

The value of $\upsilon_{}$ is extremely small, even below typical values of the sound velocity in solids and only six times higher than the sound velocity in cork~\cite{lide1990}. For a band dispersing from the zone center to the $L$ point with the same velocity, this corresponds to a band width of only 12 meV. Since entropy and specific heat of 3D Dirac fermions are proportional to $T^3/\upsilon_{}^3$ (Appendix, Eqs.~\ref{eq:cW}-\ref{eq:sW}), one would expect that such a low velocity causes the release of an enormous amount of entropy when the temperature is raised, even exceeding the acoustic phonon contribution. Since the optical conductivity of Dirac electrons is proportional to $\omega/\upsilon_{}$ (Appendix, Eq.~\ref{eq:sigslope}), we should also observe a high, linear in frequency, infrared conductivity. In Fig.~2 we compare these estimates with the experimentally measured $S(T)$ and $\sigma_1(\omega)$, and we see that the experimental values fall an order of magnitude below the calculation for $\upsilon_{}=3.1$ km/s (brown dashed line). The interband optical conductivity data at 16 K comes closest to a linear frequency dependence and can be fitted with $\upsilon_{}=41 $ km/s (green dash-dotted line). This extrapolates to 150 meV at the zone boundary and falls in the range of the 100 to 200 meV band dispersion calculated with LDA+U~\cite{wan2011,witczak-krempa2012}. 
Regarding the specific heat data, {\color{black} a quantitative comparison of the contributions from phonons, localized Nd $4f$ electrons and itinerant Ir $5d$ electrons shows that (i) below 15 K the specific heat is dominated by the localized Nd $4f$ electrons, (ii) above 15 K the specific heat is dominated by the phonons, and (iii) at all temperatures the contribution from the itinerant Ir~$5d$ electrons is negligible.} Consequently an entropy of Weyl fermions with $\upsilon_{}=3.1$ km/s as suggested by the result for $\omega_p(T)^2$ shown in Fig.~2{\bf a}, would far exceed the experimentally obtained entropy. The present data and those of Ref.~\onlinecite{ueda2012} show a perfect match above $T_N$ but, possibly due to the different types of samples ({\em e.g.} single crystal versus polycrystal), Schottky anomaly, N\'eel temperature, and specific heat jump at $T_N$ are significantly different. {\color{black}These differences are interrelated: $T_N$ being higher and the Schottky anomaly occurring at a lower temperature, together make that the jump at $T_N$ vanishes against the phonon background (see Appendix).}

This leaves us with a strange conundrum: In a single-electron model the $T^2$ temperature dependence of the free carrier spectral weight is due to thermal activation of free carriers. However, the specific heat data does not show the corresponding $S = - k_B \sum_k \left[f_k\ln f_k + (1- f_k)\ln (1- f_k) \right]$ entropy release. We therefore infer that the observed $\omega_p^2 = a T^2$ temperature dependence does {\em not} {reflect the statistical properties of non-interacting electrons}. This state of affairs calls for a radically different interpretation of the experimental data to which we return below. 

We now turn to the $ \sigma_{DC} (T)$ data, showing a striking linear temperature dependence between 15 and 37~K with a finite zero temperature intercept. This behavior of $\sigma_{DC}(T)$ has been predicted by Hosur {\em et al.} for Weyl nodes in the presence of the Coulomb interaction~\cite{hosur2012}. {\color{black}The conductivity is described by the parallel shunting of doped carriers having spectral weight $\omega_0^2$ and lifetime $\tau_0$ and thermally activated carriers with spectral weight $\omega_{p,th}^2$ proportional to $T^2$. The Drude relation $4\pi\sigma=\omega_0^2\tau_0+\omega_{p,th}^2\tau$ then implies that the momentum scattering rate of the thermally activated carriers is 
\begin{equation}
\hbar/\tau = A k_B /  T 
\nonumber
\end{equation} }
with $A=0.23$. Quite generally the ``Planckian"~\cite{zaanen2019} dissipation described by the above equation has been identified as a hallmark feature of the quantum critical state signaling the breakdown of the Fermi-liquid.
For $T<13$ K the DC conductivity drops {\em below} the linear temperature dependence, while leaving unaffected the $T^2$ dependence of  $\omega_p^2$. {It is peculiar that, despite a manifest change of behavior of $\sigma_{DC}(T)$ at 13 K, the free charge spectral weight follows a smooth $T^2$ dependence in the entire temperature range from 6 to 25 K beyond which it becomes too strongly mixed with the interband transitions. This implies that the sudden drop of  $\sigma_{DC}(T)$ 
{below 13 K} 
is caused by some form of localization of the charge carriers, with no discernable influence on the spectral weight $\omega_p^2$. 
This behavior signals a cross-over toward a state of matter characterized by an increased momentum transfer scattering. Since this overlaps with the temperature range of the Schottky anomaly of the Nd ions in the specific heat (Fig.1), it is possible that the observed scattering has -at least in part- to do with scattering by the Nd magnetic moments, as was suggested by Ueda~\cite{ueda2012}. Scattering of the itinerant Ir $5d$ electrons from the rare earth moments was also not observed in Pr$_2$Ir$_2$O$_7$, a related pyrochlore Ln$_2$Ir$_2$O$_{7}$ system that does {\em not} order at low $T$~\cite{cheng2017}. The observed Schottky anomaly can be fitted to a two-level system at each Nd site having a high temperature limiting $k_B \ln(2)$ entropy with ground-state and first excited state separated by $\Delta_0 = 6.5$ K  (Fig.1). Since this is much smaller than the 26 meV~\cite{watahiki2011} crystal field splitting between the two lowest doublets of the the $J=9/2$ manifold, the splitting $\Delta_0 = 6.5$ K signals the lifting of the degeneracy within the ground-state doublet. This requires a static exchange field at the Nd sites and is a clear indication that the time-reversal symmetry is broken. 
The freezing of Nd moments may foster opening the Mott gap of the $5d$ bands~\cite{tian2016}, the exchange fields then open the charge gap in Nd$_2$Ir$_2$O$_7$. 
}

\begin{figure}[!t]
\begin{center}
\includegraphics[width=\columnwidth]{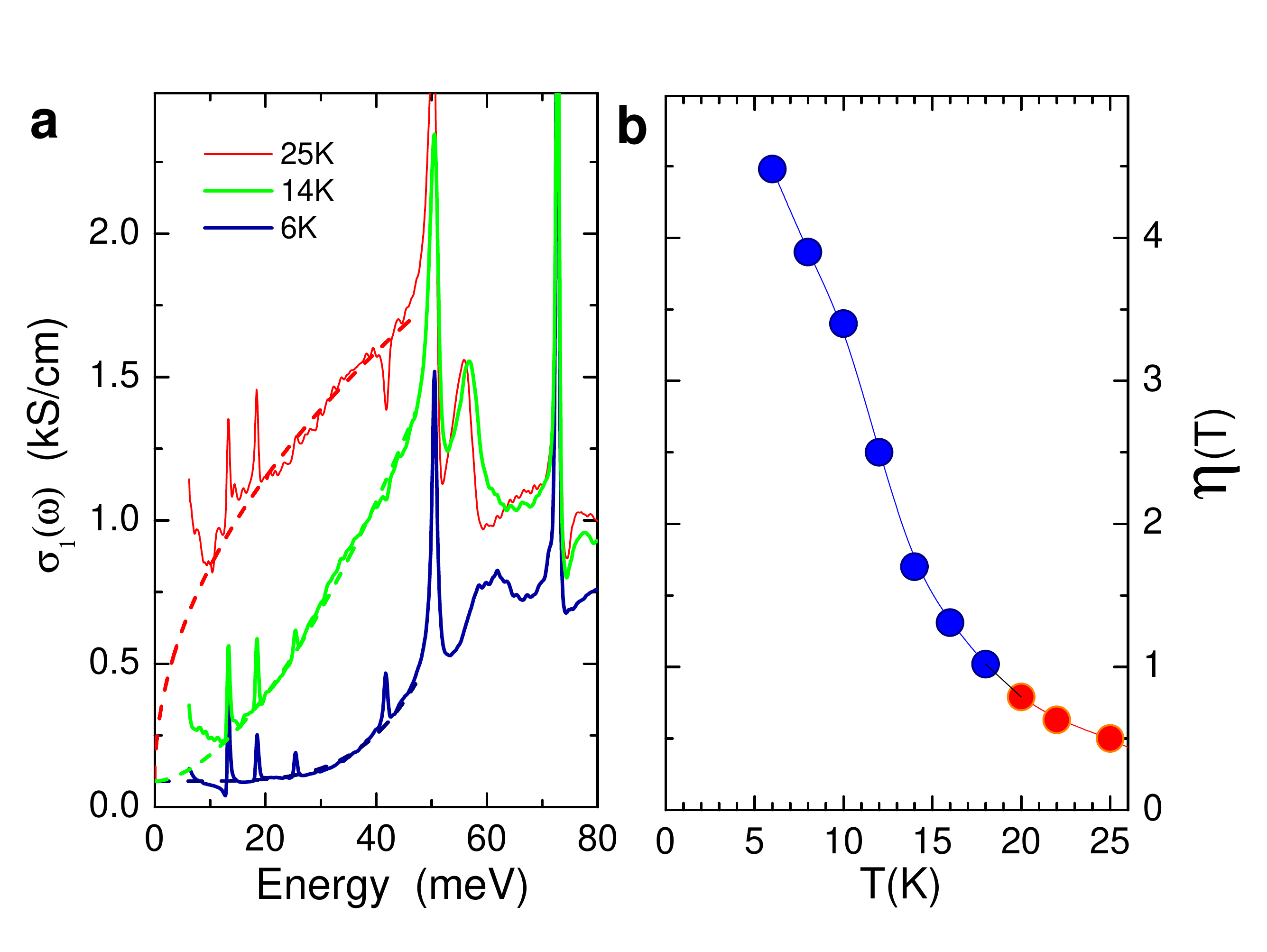}
\caption{ 
{\bf Power law analysis of the interband transitions of Nd$_2$Ir$_2$O$_{7}$.}
{\bf a}, Optical conductivity of Nd$_2$Ir$_2$O$_{7}$ with fits to the power law form $\sigma_1(\omega)=\sigma_0+a\omega^{\eta}$ with $\sigma_0=90$ S/cm. {\bf b,} Temperature dependence of the exponent $\eta$. The blue (red) color coding of the circles marks temperature regions where $\sigma_1(\omega)$ is super (sub) linear. 
}
\end{center}
\end{figure}
{We observe in Figs. 1 and 2 that the interband optical conductivity below 80 meV is strongly temperature dependent for the entire temperature range below 37~K down to the lowest measured temperature of  6 K. For most temperatures the spectral shape corresponds neither to  $\sigma_1(\omega)\propto \omega^{0.5}$ (3D quadratic band touching) nor to $\sigma_1(\omega)\propto \omega$ (3D Dirac cone). Instead the evolution as a function of temperature is phenomenologically described by the gradual evolution with temperature of a general power law  $\sigma_1(\omega)\propto \omega^{\eta}$ as shown in  Fig.~3. The temperature dependence of $\eta(T)$ reveals a gradual and strong increase from
{0.5 at 25 K} 
to 4.5 at 6 K, with an inflection point around 13 K. This evolution of $\eta(T)$ with temperature corroborates what is already obvious from direct inspection of the $\sigma_1(\omega)$ spectra, namely that upon cooling below 37~K a depletion takes place of the low energy interband transitions, consistent with the gradual (second order) transition into the Weyl semimetal phase. The spectral weight removed from low frequencies, rather than being recovered at energies directly above the gap, is transferred to a broad energy range above 1 eV. We attribute this anomalous transfer to the ``Mottness" of this system, {\em i.e.} a sensitivity to local interactions and proximity to a Mott insulating phase~\cite{phillips2006,pesin2010}.
Interestingly, the $\eta$ in Fig.~3 remains in the region between 0.5 and 1 when cooling from $T_N = 37$ K to 18 K. The inflection point around 13 K signals the appearance of the correlation-gapped Weyl semimetal phase (or ``Weyl Mott insulator"~\cite{morimoto2016}).
}

\section{Discussion}
{\color{black}
The observation of a $T^2$ zero energy spectral weight in the optical conductivity without an appreciable entropy contribution is difficult to understand. To illustrate that no general one-to-one relation exists between the entropy of the free carriers and their spectral weight we consider a non-interacting model of a semimetal for which at $T=0$ the bands touch at $E_F$ so that free carrier spectral and entropy are both zero. If, by tuning the crystal structure at $T=0$, the bands start to overlap, the free carrier spectral weight becomes finite whereas the entropy remains zero. 
{\color{black}A priori we can not rule out such a change of electronic structure as a function of temperature driven by magnetic ordering or correlation. However, since}
there are no indications in the published bandstructure of Nd$_2$Ir$_2$O$_7$~\cite{wan2011,witczak-krempa2012,wang2019} that at $T=0$ the electron and hole pockets should touch at $E_F$, we infer that the free carrier optical conductivity of Nd$_2$Ir$_2$O$_7$ is collective, and that due to limitations on the number of collective degrees of freedom the associated entropy is small. This state of affairs is present in superconductors which are known to be heat insulators while being perfect charge conductors. Another case is presented by charge density waves where collective current transport can be accommodated by sliding of the charge density wave~\cite{gruner1988}. 
However neither the superconducting nor the CDW scenario is compatible with the observed $T^2$ dependence of the zero energy spectral weight. 
Recently Morimoto and Nagaosa~\cite{morimoto2016} (hereafter MN) predicted that a Weyl-metal with a non-local electron-electron interaction in the limit of forward scattering becomes a gapped Weyl Mott-insulator. Due to thermal activation across the insulator gap one expects the intensity of the Drude peak and of the entropy to be non-zero and that both of them follow an Arrhenius-type temperature dependence - which is incompatible with our data. However, this model also predicts a gapless continuum of collective (electron-hole) excitations at $q=0$ and MN conjecture that for realistic interactions this turns into a collective mode with a linear dispersion. Such a generalization would be extremely interesting in view of its consequences for the optical conductivity and the entropy.

\section{Conclusions}
We observed a $T^2$ temperature dependence of the weight of the zero energy mode in Nd$_2$Ir$_2$O$_7$ without the corresponding $T^3$ contribution to the entropy. This poses a challenge to single particle band-structure approaches, as it indicates that the collective sector and the single particle sector are effectively decoupled. 
\section{Acknowledgements}
DvdM acknowledges insightful discussions with Dmitry Abanin and Naoto Nagaosa. 
This project was supported by the Swiss National Science Foundation (Project No. 200020-179157). This work is partially supported by CREST(JPMJCR18T3), Japan Science and Technology Agency (JST), by Grants-in-Aids for Scientific Research on Innovative Areas (15H05882 and 15H05883) from the Ministry of Education, Culture, Sports, Science, and Technology of Japan, and by Grants-in-Aid for Scientific Research (19H00650). Work at JHU was supported through the Institute for Quantum Matter, an EFRC funded by the U.S. DOE, Office of BES under DE-SC0019331.
%

\appendix


\section{Experiments} 
%
\noindent{\bf Sample synthesis.} 
Single crystals of Nd$_2$Ir$_2$O$_7$ were grown by the same flux method using polycrystalline powder and KF flux as described in Ref.~\onlinecite{ishikawa2012}. The crystal used for this study had a weight of 1.83~mg, it was  0.4~mm thick and it had a trapezium shaped 0.5~mm$^2$ surface area that was used for the optical experiments.

\noindent{\bf Magnetic susceptibility.}
Zero field cooled (ZFC) and field cooled (FC) magnetization was measured as a function of temperature using the vibrating sample mode (VSM) of a superconducting quantum interference device SQUID magnetometer (Quantum Design MPMS$\textregistered$).  The VSM frequency was set to 14~Hz. For the ZFC measurement, the sample was cooled from room temperature down to 4~K without any applied field.
At low temperature, a magnetic field of 1000~Oe was applied and the magnetization was measured upon warming at a constant rate of 0.45~K/minute. For the FC measurement, the sample was cooled in a field and the data were collected in the same conditions as for ZFC.

\noindent
The susceptibility of a  system of $n$ paramagnetic ions per unit volume is given by the Curie law
\begin{equation}
\chi(T) =  \frac{n p^2 \mu_B^2 }{3 k_B T} 
\end{equation}
where  $p$ is the effective moment. Vice versa, if there is no coupling between the moments the effective moment $p(T)$ defined as 
\begin{equation}
p(T) \equiv  \sqrt{\frac{3 k_B T \chi(T)}{ n \mu_B^{2}}}
\label{eq:p}
\end{equation}
should be independent of temperature. The quantity $p(T)^2$  is displayed in Fig.~1{\bf a}.

\noindent{\bf DC transport.}
The DC resistivity was measured in the temperature range from 1.8 to 300~K using a Quantum Design Physical Property Measurement System (PPMS$\textregistered$) in 4-point geometry with an excitation current of 0.5~mA. Electric contacts on the samples were done using silver paste.
The resistivity was calculated from the resistance using sample thickness, width and the distance between the contacts. Comparison with the optical $\sigma(\omega,T)$ extrapolated to $\omega=0$ showed a lower value of the conductivity measured with the PPMS$\textregistered$ at all temperatures. Since the contact layout on a small sample may result in an overall scaling of the resistivity, and since the optical conductivity at 300~K is frequency independent in the far infrared, we have used the optical $\sigma(\omega,300~\mbox{K})$  (see next section) extrapolated to $\omega=0$ to calibrate the DC resistivity shown in Fig.~1{\bf a} of the main text.

\begin{figure}[!h!!]
\begin{center}
\includegraphics[width=\columnwidth]{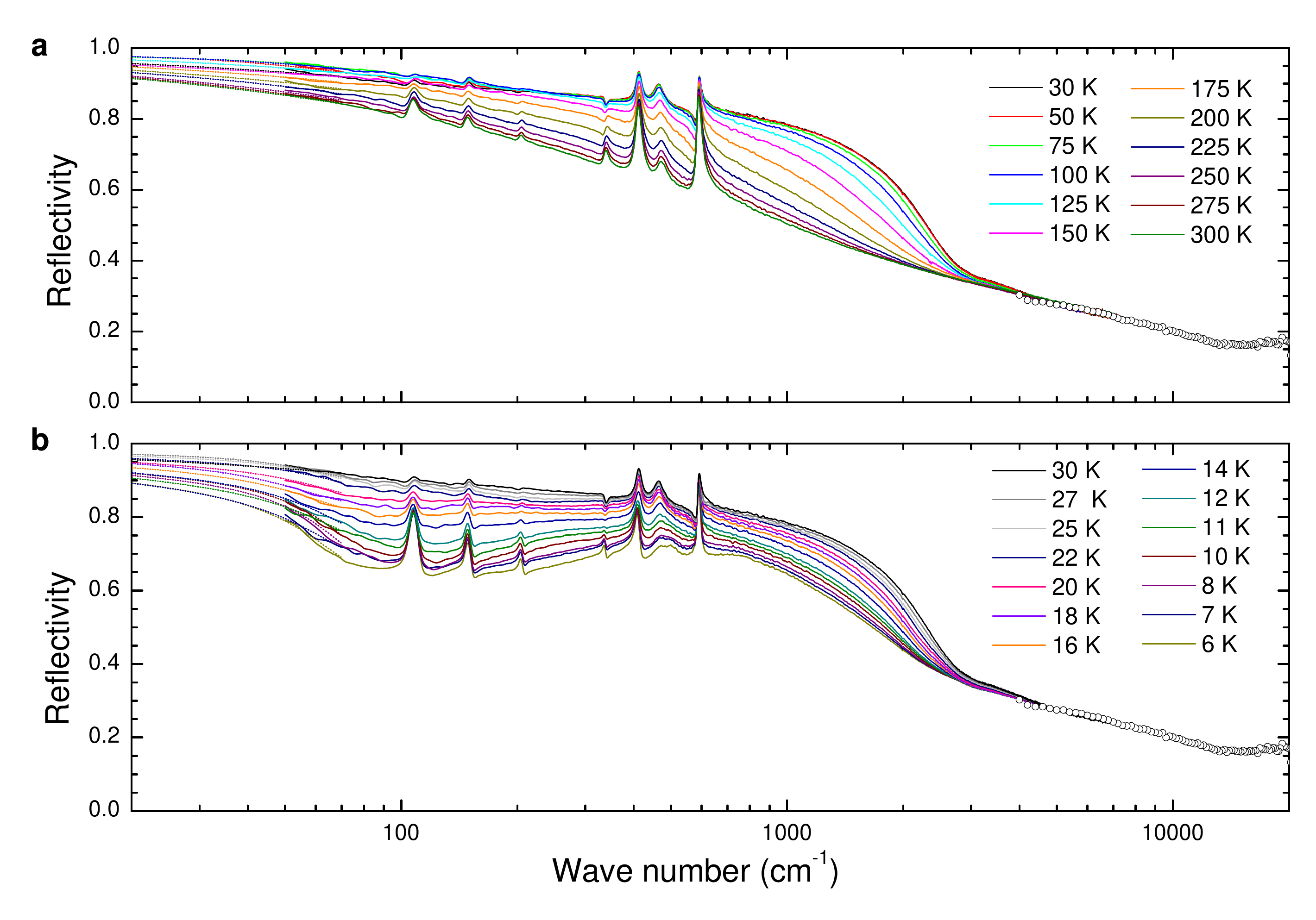}
\caption{ 
{\bf Reflectance spectra}. 
Solid curves: Near normal incidence reflectivity, $R=|r|^2$, of Nd$_2$Ir$_2$O$_{7}$ for selected temperatures.
Dotted curves below 50 cm$^{-1}$: Extrapolations using simultaneous Drude-Lorentz fitting to the reflectance spectra and the ellipsometric data between 4000 and 18000 cm$^{-1}$ of Fig. 5.
{\bf a}, Between 30 and 300 K.
{\bf b}, Between 6 and 30 K.
}
\end{center}
\end{figure}
%

\noindent{\bf Optical measurements and data analysis.}
We measured the near normal reflectivity $|r(\omega)|^2$ from {\{5~meV ; 1.5~eV\}} with a Fourier transform spectrometer combined with a UHV flow cryostat, using in-situ gold evaporation for calibrating the signal. These data are displayed in Fig.~4. In the energy range {\{0.5~eV ; 3~eV\}} we measured the complex dielectric function  $\epsilon(\omega)=\epsilon_1(\omega)+i\epsilon_2(\omega)$ using  ellipsometry of the ab-plane of our samples at an incident angle 65$^\circ$ relative to the normal. The result is displayed in Fig.~5. From $\epsilon(\omega)$ we calculated amplitude $|r|$ and phase ${\phi}$ of the normal incidence reflectivity using the Fresnel equation, providing an excellent match with the aforementioned reflectivity data. Fitting $|r(\omega)|$ and $\phi(\omega)$ simultaneously with a Drude-Lorentz expansion of $\epsilon(\omega)$ provided extrapolations of $|r(\omega)|$ in the ranges {\{0; 5~meV\}} and \{3~eV ; $\infty$\}. Application of the  Kramers-Kronig relation to $\ln(|r(\omega)|)$ in the range \{0; $\infty$\} provided the phase spectrum $\phi(\omega)$ at all frequencies, in particular in the range {\{5~meV ; 0.5~eV\}} where the phase was not measured using ellipsometry. The complex dielectric function $\epsilon(\omega)$ and the optical conductivity $4\pi\sigma_1(\omega)=\omega\epsilon_2(\omega)$ were calculated in the entire range of the experimental data using the Fresnel equation for the reflection coefficient.
The effective particle number in the primitive cell (2 formula units, 4 Ir ions) of volume $V_{pc}=2.79\cdot 10^{-22}$~cm$^{3}$ was calculated using the expression
\begin{equation}
N_{eff}(\omega)=\frac{2m_e V_{pc}}{\pi e^2}\int_0^{\omega}\sigma_1(\omega')d\omega'
\label{eq:neff}
\end{equation}
\begin{figure}[!h!!]
\begin{center}
\includegraphics[width=0.6\linewidth]{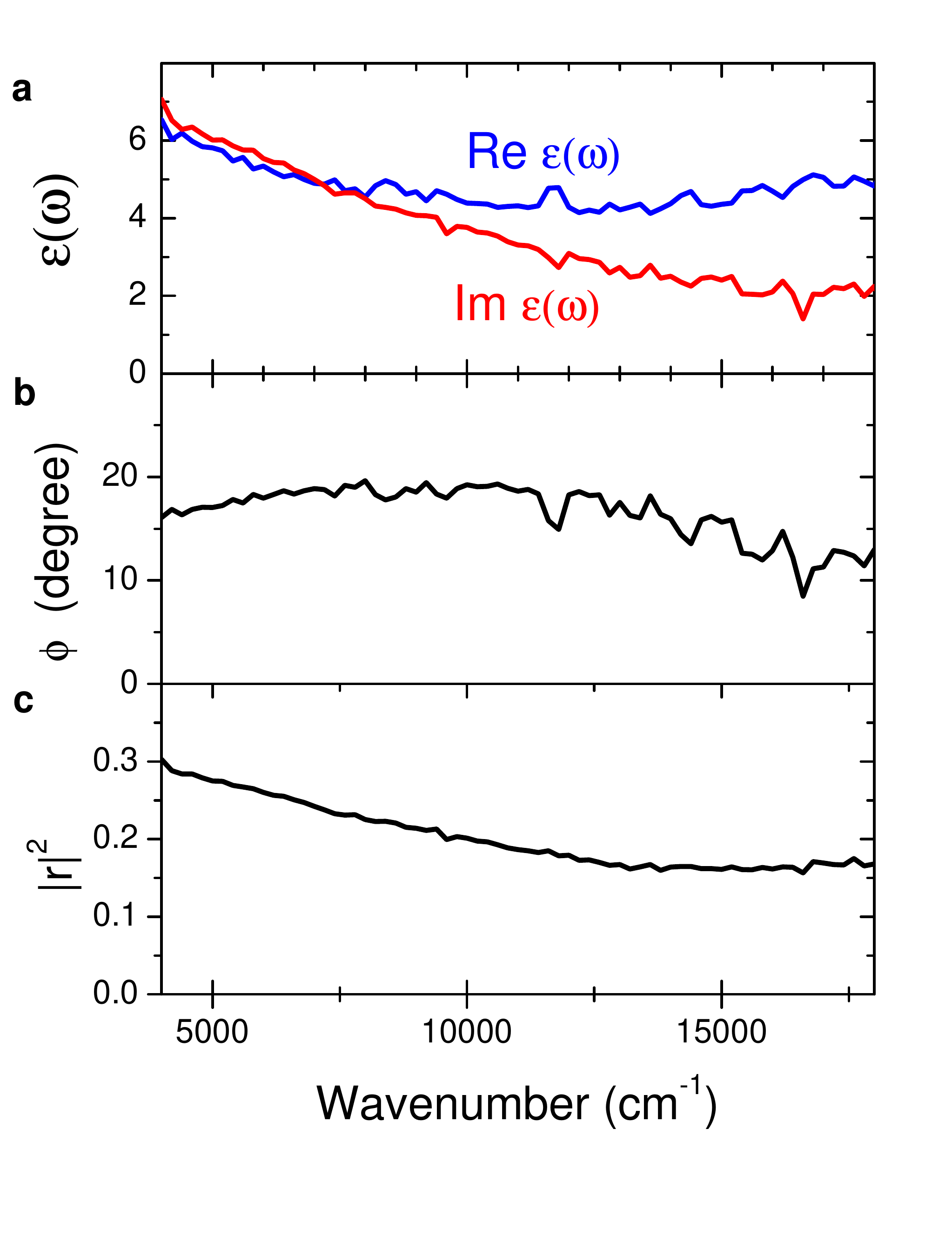}
\caption{ 
{\bf Ellipsometric data between 4000 and 18000 cm$^{-1}$.}
{\bf a}, Real and imaginary part of the dielectric function at room temperature measured using spectroscopic ellipsometry at $\theta = 65^0$ with the surface normal. 
{\bf b}, Phase of the normal incidence reflection coefficient using the Fresnel equation $|r|e^{i\phi}=(1-\sqrt{\epsilon})/(1+\sqrt{\epsilon})$.
{\bf c}, Absolute square of the reflection coefficient.
}
\end{center}
\end{figure}

\noindent{\bf Specific heat experiments.}
The specific heat of the $Nd_{2}Ir_{2}O_{7}$ single crystal was obtained in a Quantum Design Physical Property Measurement System (PPMS$\textregistered$) with the heat capacity relaxation technique from 5~K to 55~K in increments of 1 K with a temperature relaxation amplitude of 1\% of the base temperature ({\em e.g.} at 10~K the temperature relaxation amplitude is about 0.1~K).
We first measured the empty heat capacity puck with a small amount of Apiezon N grease (used for glueing for the sample with a good thermal contact) in order to extract the heat capacity of the platform + Apiezon N grease, $C_{add}(T)$. We then added the sample and measured the total heat capacity $C_{tot}(T)$. This provided the heat capacity of the sample $C_{sample}(T) = C_{tot}(T) - C_{add}(T)$. Using the mass of the sample $M = 1.830 \pm 0.005$~mg and the molar mass per $Nd_{2}Ir_{2}O_{7}$ formula unit $M_{fu} = 784.92$~g/mol we calculated the molar specific heat $C(T)=C_{sample}(T) M_{fu}/M$. Since the primitive cell contains two formula units of $Nd_{2}Ir_{2}O_{7}$, the specific heat per primitive cell is, in units of $k_B$: 
$c(T)=2C(T)/R$ where $R$ is the gas constant. 
The entropy per primitive cell (in units of $k_B$) is calculated from the specific heat using:
\begin{eqnarray}
s(T) = \int_0^T\frac{c_{}(T')}{T'} dT'
\label{eq:S}
\end{eqnarray}
%

\noindent
The experimental specific heat data in the range from 5 to 40 K (Fig.~1{\bf b} of the main text) was fitted to the expression (Eqs.~\ref{eq:Debye} and~\ref{eq:schottky_c})
\begin{equation}
c(T) = c_{f}(T)+ c_{v}(T)
\label{eq:CV}
\end{equation}

\noindent
We have fitted the experimental data from 5 to 40~K fixing $T_N=37$~K and $\eta=2$. This provided the parameters $\Delta_{0}$, $a_3$ and $a_5$.

\section{Theoretical expressions}
\noindent{\bf Phonon contribution to the specific heat.}
To model the vibrational contribution to the specific heat we use 
\begin{eqnarray}
c_{v}(T)=a_3T^3 - a_5T^5
\label{eq:Debye}
\end{eqnarray}
\noindent
The parameters $a_3$ and $a_5$ effectively subsume -for the relevant range of temperatures- the thermodynamic properties of all (3 acoustic and 63 optical) vibrational modes in the primitive cell.

\noindent{\bf Magnetic fluctuation contribution to the specific heat.}
We assume that the Ir sub-lattice induces an exchange field at the Nd-site causing the ground state doublet to split with an energy difference $\Delta_{}$. The entropy per primitive cell is, in units of $k_B$
\begin{equation}
s_{f} =  4\left[  
\frac{\ln\left(1+e^{-\Delta_{}/T} \right)}{1+e^{-\Delta_{}/T}}
+
\frac{\ln\left(1+e^{\Delta_{}/T} \right)}{1+e^{\Delta_{}/T}}
\right]
\label{eq:schottky_s}
\end{equation}
where the factor of $4$ at the righthand side accounts for the fact that each primitive cell contains $4$ Nd$^{3+}$-ions. 
We assume $\Delta_{}=0$ for $T>T_N$ and below $T_N$ we adopt the following approximation for the temperature dependence
\begin{equation}
\Delta_{} = {\Delta_{0}}\sqrt {1 - {{\left( {\frac{T}{{{T_N}}}} \right)}^\eta }} 
\label{eq:J}
\end{equation}
The relation $c_{f}(T)=T ds_{f}(T)/dT$ then gives for the specific heat per primitive cell:
\begin{equation}
c_{f}(T) = 4
\frac{e^{\Delta_{}/T}}{\left( e^{\Delta_{}/T}+1 \right)^2}\left[ {\frac{{{\Delta_{}^2}}}{{{T^2}}} + \eta \frac{{\Delta_{0}^2}}{2}\frac{{{T^{\eta  - 2}}}}{{T_N^{^\eta }}}} \right]
\label{eq:schottky_c}
\end{equation}

\noindent{\bf Itinerant Ir $5d$ electron contribution to the specific heat.}
Specific heat and entropy per primitive cell corresponding to $g$ Weyl nodes are, in units of $k_B$
\begin{eqnarray}
c_W(T) &=& g V_{pc} \left(\frac{k_BT}{ \hbar \upsilon_{}}\right)^3 \frac{4}{ \pi^2}  \int\limits_0^\infty  \frac{x^3}{e^x+1}  dx  
\label{eq:cW}
\\
s_W(T) &=& \frac{g}{3} V_{pc} \left(\frac{k_BT}{ \hbar \upsilon_{}}\right)^3 \frac{4}{ \pi^2}  \int\limits_0^\infty  \frac{x^3}{e^x+1}  dx  
\label{eq:sW}
\end{eqnarray}
where $V_{pc}$ is the volume of the primitive cell (two formula units of Nd$_2$Ir$_2$O$_7$).
\noindent
For the electronic contribution of the Ir~$5d$ states for $T>T_N$ we use the specific heat of a single quadratic band touching point
\begin{eqnarray}
c_{QBT}(T) &=&  \frac{5\sqrt{2}}{\pi^2} \frac{m^{3/2} V_{pc}}{\hbar^3} \left(k_BT\right)^{3/2}   \int\limits_0^\infty  \frac{x^{3/2}}{e^x+1}  
\label{eq:cQBT}
\end{eqnarray}
We use  $m\sim 6.3 m_0$ where $m_0$ is the free electron mass, which is obtained from fitting ARPES data of Pr$_2$Ir$_2$O$_7$ (see Refs.~15,16 of the main text), a compound that is paramagnetic at all temperatures.
To describe the second order transition from Weyl semimetal to paramagnetic QBT semimetal we use for $T\le T_N$
\begin{eqnarray}
s(T)&=& s_W(T) +\left(T/T_N\right)^{\eta}  \left[s_{QBT}(T)- s_W(T))\right]
\label{eq:sQBT}
\end{eqnarray}
so that $ s(0)=s_W(0)$ and $s(T_N)=s_{QBT}(T_N)$.  For $s_W(T)$ we use Eq.~\ref{eq:sW}, taking $\upsilon=41$ km/s and $g=24$. In Fig.~7 we use $\eta=2$. Taking a larger value, {\em e.g.} $\eta=4$, does not significantly  change the outcome. The entropy described by Eq.~\ref{eq:sQBT} is continuous at all temperatures, and it has a change of slope at $T_N$ corresponding to a (negative) jump in the specific heat $c=Tds/dT$ at $T_N$.

\noindent{\bf Optical properties of Weyl nodes.}
We summarize the expressions derived by Tabert {\em et al.}$^{31,32}$, which we use in the discussion of the paper.
The $T=0$ optical conductivity of a 3D Dirac semimetal is
\begin{eqnarray}
\sigma_1(\omega) &=& \frac{\alpha }{ 24 \pi} \frac{g c }{ \upsilon_{}} \omega 
\label{eq:sigslope}
\end{eqnarray}
where $\alpha=1/137$ the fine structure constant, $g$ the number of Dirac cones, $c$ the light velocity and $\upsilon_{}$ the Dirac velocity. 
The temperature dependence of the free carrier spectral weight is
\begin{eqnarray}
\omega_p(T)^2&=&\omega_p(0)^2+\frac{2\pi \alpha}{9} \frac{g c }{\upsilon_{}}  \left(k_BT\right)^2  
\label{eq:wp2}
\end{eqnarray}
Moving the chemical potential $\mu$ away from the Dirac point results in a carrier density 
\begin{eqnarray}
n_0= \frac{g\mu^3 }{6 \pi^2 \hbar^3 \upsilon_{}^3 }
\label{eq:n0}
\end{eqnarray}
and a residual Drude spectral weight 
\begin{eqnarray}
\omega_p(0)^2= \left(\frac{32\pi g n_0^2 }{ 3 }\right)^{1/3}  \alpha c  \upsilon_{}
\label{eq:omega02}
\end{eqnarray} 
\section{Data availability.}
The datasets of this study are available in Ref.~\cite{yareta}.  All other data that support the plots within this paper and other findings of this study are available from the corresponding author upon reasonable request.
\bibliographystyle{naturemag}
\bibliography{biblio}
\end{document}